\documentstyle[12pt]{article}
\newcommand{\be}{\begin{equation}}
\newcommand{\ee}{\end{equation}}
\newcommand{\ba}{\begin{eqnarray}}
\newcommand{\ea}{\end{eqnarray}}

\begin{document}
\begin{center}
 {\bf\Large{
   Surface terms, angular momentum and  Hamilton - Jacobi formalism  }}
\end{center}
\begin{center} {\bf
Yurdahan G\"uler}\footnote[1]{E-Mail:~~yurdahan@cankaya.edu.tr},
{\bf Dumitru Baleanu}\footnote[2]{ On leave of absence from
Institute of Space Sciences, P.O BOX, MG-23, R 76900
Magurele-Bucharest, Romania, E-mail: dumitru@cankaya.edu.tr}, {\bf
Murat Cenk}\footnote[3]{E-Mail:~~mcenk@cankaya.edu.tr}
\end{center}
\begin{center}
Department of Mathematics and Computer Sciences, Faculty of Arts
and Sciences, Cankaya University-06530, Ankara , Turkey
\end{center}
\begin{abstract}
Quadratic Lagrangians are introduced adding surface terms to a
free particle Lagrangian. Geodesic equations are used in the
context of the Hamilton-Jacobi formulation of constrained sysytem.
Manifold structure induced by the quadratic Lagrangian is
investigated.

\end{abstract}

\section{Introduction}

 There are mainly two basic methods to investigate constrained
systems. First one is the method initiated by Dirac \cite{Dirac}.
This method is based upon the classification of constraints and
consistency conditions derived from the the variations of the
constraints. The symplectic structure is established by defining
Dirac brackets. The second method is the treatment of the
constrained systems by Caratheodory's equivalent Lagrangian method
\cite{car},\cite{g5}, \cite{g6}. Since this method leads us to the
Hamilton-Jacobi equations it will be called Hamilton-Jacobi(HJ)
approach. Equivalent Lagrangian method \cite{car} leads us to a
set of Hamilton-Jacobi equations \cite{g5}, \cite{g6}. Despite of
many attempts to clarify the many aspects of Hamilton-Jacobi
formalism (HJ) we still have some subtle ones. The main difference
between Dirac's formulation \cite{Dirac} and (HJ) approach
\cite{gb1}, \cite{gb2}, \cite{gb3} is the fact that the second one
is multi Hamiltonian approach depending on the rank of the Hessian
matrix.
 On the other hand the boundary conditions
\cite{vergara},\cite{montesino}, \cite{jabbari}, \cite{devecchi}
play an important role for constrained systems \cite{const} and
they are crucial for (HJ) formulation \cite{car},\cite{bagu}. The
presence of many Hamiltonians, which are constraints having a
special forms, suggests us to investigate the possibility to
describe the superintegrable systems \cite{superi} and to obtain
some integrable geometries.

 In this paper we start with
a free Lagrangian theory  and we add a surface term involving the
components of the angular momentum. By a suitable choosing of a
Lagrange multipliers we obtain a quadratic Lagrangian which can be
singular or not. The aim of this paper is to study quadratic,
singular Lagrangians, using the (HJ) formulation. This paper is
arranged as follows: In Sec. 2 (HJ) formalism is briefly
explained. In Sec. 3 the method of obtaining quadratic Lagrangians
is presented and two non-singular quadratic systems are studied.
In Sec. 4 the singular case is analyzed using the (HJ) formalism.
Conclusions are presented in Sec. 5. In Annex some geometrical
properties of the obtained metrics are shown.

\section{Hamilton-Jacobi formalism}

 Let us assume that the Lagrangian L is singular and the Hessian
matrix
$\frac{\partial^{2}L}{\partial\dot{q^a}\partial\dot{q^b}},\\
a,b=1,2,..,n-r$, has rank n-r.The canonical Hamiltonian  $H_{0}$
is defined as
$$H_{0}=p_{i}\dot{q^{i}}-L(t,q^{i},\dot{q^{i}})$$
The other Hamiltonians are obtained from the definitions $
p_{\alpha}=\frac{\partial L}{\partial \dot{q^{\alpha}}}\mid_{{\dot
q_a}=\omega_a}$ and they have the forms
\begin{equation}
H^{'}_{\alpha}=H_{\alpha}(t_{\beta},q_{a},p_{a})+p_{\alpha}
\end{equation}
where $\alpha,\beta=n-r+1,..,n;\quad a=1,2,..,n-r$.
 Thus, one arrives at the following set of
Hamilton-Jacobi partial differential equations
\begin{equation}
\frac{\partial S}{\partial
q^{\alpha}}+H_{\alpha}(t^{\beta},q^{a},p_{a}=\frac{\partial
S}{\partial q^{a}})=0
\end{equation}

The following total differential equations lead us to determine\\
the Hamilton-Jacobi function $S(t^{\alpha},q^{a})$:
\begin{eqnarray}\label{sys}
  dq^{a} &=& \frac{\partial H_{\alpha}^{'}}{\partial p_{a}}dt^{\alpha} \\
  dp_{a} &=& -\frac{\partial H_{\alpha}^{'}}{\partial q^{a}}dt^{\alpha}  \\
  dp_{\mu}&=&-\frac{\partial H_{\alpha}^{'}}{\partial t^{\mu}}dt^{\alpha}
\end{eqnarray}
Solutions of these equations $S(q_i)$ as \be
dz=(-H_{\alpha}+p_{a}\frac{\partial H_{\alpha}^{'}}{\partial
p_{a}})dt^{\alpha}, \ee where $z=S(t^{\alpha},q^{a})$. Since the
equations of motion are total differential equations, one should
check the differentiability conditions, consistency conditions in
Dirac formulation. It is proved that the system of equations
(\ref{sys}) is differentiable if and only if the following Poisson
brackets equations are satisfied :
\begin{equation}
    [H_{\alpha}^{'},H_{\beta}^{'} ]=0.
\end{equation}

In general some of these Poisson brackets may not vanish
identically. In such cases one should include the new functions as
''new'' constraints. Poisson brackets of these constraints with
$H_{\alpha}^{'}$ should be considered also. This procedure should
continue until all Poisson brackets vanish identically. This
approach was studied in detail in references \cite{g5},\cite{g6}.

\section{The method}

Let us assume that a given Lagrangian $L({\dot q^i},q^i)=
\frac{1}{2}a_{ij}{\dot q^{i}}{\dot q^{j}}$ admits a set of
independent constants of motion denoted by $L_{i}, i=1,\cdots n$.
Our aim is to treat the new Lagrangian $L^{'}=
L+{\dot\lambda^i}L_i$ as a quadratic Lagrangian expresses as
\begin{equation}\label{mu}
L^{'}=\frac{1}{2}g_{ij}\dot{q^{i}}\dot{q^{j}}.
\end{equation}
Let us assume that the matrix $g_{ij}$ is non-singular.

 Treating
$g_{ij}$ as the ''metric tensor'' of a manifold the geodesic
equations

\begin{equation}
    \ddot{q^{a}}=\frac{-1}{2}g^{ab}(\frac{\partial
g_{bm}}{\partial q^{k}}+\frac{\partial g_{bk}}{\partial
q^{m}}-\frac{\partial g_{mk}}{\partial
q^{b}})\dot{q^{k}}\dot{q^{m}}
\end{equation}
are equivalent to the Euler-Lagrange equations of (\ref{mu}). Here
$g^{ab}$ is the inverse of $g_{ab}$.
 In this formulation the canonical Hamiltonian
of the system is expressed as $H=\frac{1}{2}g^{ab}p_{a}p_{b}$. In
this way the motion of a particle is described by geodesic
equations of a Riemannian manifold with metric tensor $g_{ij}$.
The above motion is treated inside of (HJ) formalism and the
geometrical properties of the obtained metrics $g_{ij}$ will be
investigated. In the following we will add to a free Lagrangian
the components of the angular momentum.

 {\bf 1.}As the first example of a non-singular quadratic system,
 let's consider the following Lagrangian
 \begin{equation}\label{priimaa}
 L=\frac{1}{2}(\dot{x}^2+\dot{y}^2)+\dot{\lambda_{3}}(x\dot{y}-y\dot{x}),
\end{equation}
which can be expressed as
 $L=\frac{1}{2}a_{ij}{\dot q^{i}}{\dot q^{j}}$, where
 \be\label{matrice}
 a_{ij}=\left( \begin{array}{ccc}
 1 & 0 &-y \\
 0 & 1& x \\
-y & x& 0 \end{array}\right). \ee This Lagrangian corresponds to
the motion of a particle moving on a plane such that $L_z$ is
constant.\\
Hamiltonian has the following form
\begin{equation}
H_c=\frac{1}{2}a^{-1}_{\mu\rho}p_{\mu}p_{\rho} \quad where \quad
\mu,\rho=1,..,3
\end{equation}
and $a^{-1}_{\mu\rho}$ is the inverse of $a_{\mu\rho}$. From
Hamiltonian we find
\begin{equation}\label{hhh}
\begin{array}{c}
dx=a_{1\rho}^{-1}p_{\rho}dt,\quad dy = a_{2\rho}^{-1}p_{\rho}dt
,\quad d\lambda_{3}=
a_{3\rho}^{-1}dt\\
dp_{x}= \frac{-1}{2}(\frac{\partial a_{\mu\rho}^{-1}}{\partial
x})p_{\mu}p_{\rho}dt, \quad dp_{y}= \frac{-1}{2}(\frac{\partial
a_{\mu\rho}^{-1}}{\partial y})p_{\mu}p_{\rho}dt, \quad
dp_{\lambda_{3}}= 0
\end{array}
\end{equation}
In order to find equation of motion we must solve (\ref{hhh}).
However we will prefer geodesic equations that produce easier
equations and equivalent to the (\ref{hhh}). The geodesic
equations corresponding to (\ref{matrice}) are
\begin{equation}\label{um}
\begin{array}{c}
    \ddot{x}=
\frac{2x\dot{\lambda_{3}}}{x^2+y^2}(x\dot{y}-y\dot{x}),\\
  \ddot{y} = \frac{2y\dot{\lambda_{3}}}{x^2+y^2}(x\dot{y}-y\dot{x}), \\
  \ddot{\lambda_{3}} =
  \frac{-2\dot{\lambda_{3}}}{x^2+y^2}(x\dot{x}+y\dot{y}).
\end{array}
\end{equation}
The aim is to solve the system of equations given by (\ref{um}).
Notice that  (\ref{um}) implies that
\begin{equation}\label{um1}
\frac{\ddot{\lambda_{3}}}{\dot{\lambda_{3}}}=\frac{-2(x\dot{x}+y\dot{y})}{x^{2}+y^{2}},
\end{equation}
which admits solutions as
\begin{equation}\label{um2}
\lambda_{3}=\int\frac{C}{x^2+y^2}dt,
\end{equation}
where C is constant. Taking into account that
\[\frac{\ddot{x}}{\ddot{y}}=\frac{x}{y} \quad and\quad denoting  \quad \frac{\ddot{x}}{x}=\frac{\ddot{y}}{y}=k \]
we divide our problem into two parts. The first one corresponds to
the case when  $k < 0$. We obtain the solutions as
\begin{equation}\label{y}
x(t)=c_{1}\sin(\sqrt{k}t)+c_{2}\cos(\sqrt{k}t),\quad
y(t)=c_{3}\sin(\sqrt{k}t)+c_{4}\cos(\sqrt{k}t).
\end{equation}
 If we impose some initial conditions one can
find specific solutions of equations.Let us assume that $c_{1}=0$,
$ c_{2}=1$, $c_{3}=1$ and $c_{4}=0$. From (\ref{y}) we find that
$x(t)=\cos(\omega t) \quad and \quad y(t)=\sin(\omega t)$ where
$\omega =\sqrt{k}$. Also from equation (\ref{um2}) we obtain that
$\lambda_{3}=Ct+B$ where B is  a constant. On the other hand
 $p_{x}$, $p_{y}$ and $p_{\lambda_{3}}$ are given as follows
\begin{equation}\label{pu}
\left(%
\begin{array}{c}
  p_{x} \\
  p_{y} \\
   p_{\lambda_{3}}
\end{array}%
\right) =
\left(%
\begin{array}{ccc}
  1 & 0 & -y \\
  0 & 1 & x \\
  -y & x & 0 \\
\end{array}%
\right)
\left(%
\begin{array}{c}
  \dot{x} \\
  \dot{y} \\
  \dot{\lambda_{3}} \\
\end{array}%
\right)
\end{equation}
If we solve (\ref{pu}) we find
\begin{equation}\label{p}
p_{x}=\sin(\omega t)(-\omega-C)\quad p_{y}=\cos(\omega
t)(\omega+C)\quad p_{\lambda_{3}}=\omega
\end{equation}

 In the case of $k>0$ the solutions are given by
\begin{equation}\label{ex}
x(t)=c_{1}e^{\sqrt{k}t}+c_{2}e^{-\sqrt{k}t},
y(t)=c_{3}e^{\sqrt{k}t}+c_{4}e^{-\sqrt{k}t},
\end{equation}
where $c_{i}, i=1,..,4$ are constants. As in the previous case
$\lambda_{3}$ is given by (\ref{um2}).Using (\ref{ex}) and
imposing $A=c_{3}c_{2}-c_{1}c_{4}\neq0$ we obtain the following
equation
$(c_{4}^{2}+c_{3}^{2})x^{2}+(c_{1}^{2}+c_{2}^{2})y^{2}-2xy(c_{2}c_{4}+c_{1}c_{3})=A^2.$

The metric from (\ref{matrice}) is conformaly flat having its
Ricci scalar $R=\frac{2}{x^2+y^2}$.

 {\bf 2.} Let us add now
two components of the angular momentum to the Lagrangian of a free
three-dimensional particle. We obtain
\begin{equation}\label{doii}
L=\frac{1}{2}(\dot{x^{2}}+\dot{y^{2}}+\dot{z^{2}})+\dot{\lambda_{1}}(y\dot{z}-z\dot{y})+\dot{\lambda_{2}}(z\dot{x}-x\dot{z}),
\end{equation}
and from (\ref{doii}) we identify the metric $a_{ij}$ as

\begin{equation}\label{cinci} a_{ij}=\left(%
\begin{array}{ccccc}
  1 & 0 & 0 & 0 & z \\
  0 & 1 & 0 & -z & 0 \\
  0 & 0 & 1 & y & -x \\
  0 & -z & y & 0 & 0 \\
  z & 0 & -x & 0 &0 \\
\end{array}\right)
\end{equation}

 In this case the geodesic equations have the following forms

\begin{equation}
\ddot{x}=\frac{x}{x^2+y^2+z^2}[2\dot{\lambda_{1}}(y\dot{z}-z\dot{y})+2\dot{\lambda_{2}}(z\dot{x}-x\dot{z})]
\end{equation}
\begin{equation}
\ddot{y}=\frac{y}{x^2+y^2+z^2}[2\dot{\lambda_{1}}(y\dot{z}-z\dot{y})+2\dot{\lambda_{2}}(z\dot{x}-x\dot{z})]
\end{equation}
\begin{equation}
\ddot{z}=\frac{z}{x^2+y^2+z^2}[2\dot{\lambda_{1}}(y\dot{z}-z\dot{y})+2\dot{\lambda_{2}}(z\dot{x}-x\dot{z})]
\end{equation}
\begin{equation}\label{l1}
\ddot{\lambda_{1}}=\frac{2}{x^2+y^2+z^2}[-\dot{\lambda_{1}}(\frac{x^2\dot{z}+z^2\dot{z}}{z}+y\dot{y})+\dot{\lambda_{2}}y(\frac{z\dot{x}-x\dot{z}}{z})]
\end{equation}
\begin{equation}\label{l2}
 \ddot{\lambda_{2}}=\frac{2}{x^2+y^2+z^2}[x\dot{\lambda_{1}}(\frac{z\dot{y}-y\dot{z}}{z}+)-\dot{\lambda_{2}}(\frac{y^2\dot{z}+z^2\dot{z}}{z}+x\dot{x})]
\end{equation}
These equations form a system of  nonlinear ordinary differential
equation and to solve this system we should notice that
$\frac{\ddot{x}}{x}=\frac{\ddot{y}}{y}=\frac{\ddot{z}}{z}$.
Assuming that the above ratio is a negative constant $k^{'}$ we
obtain
\begin{equation}
x(t)=c_{1}\sin(\sqrt{k}t)+c_{2}\cos(\sqrt{k}t)
\end{equation}
\begin{equation}
y(t)=c_{3}\sin(\sqrt{k}t)+c_{4}\cos(\sqrt{k}t)
\end{equation}
\begin{equation}
z(t)=c_{5}\sin(\sqrt{k}t)+c_{6}\cos(\sqrt{k}t),
\end{equation}
where $-k^{'}=k$ and the constants $c_i,\quad i=1...4$ are
subjected to the restriction that
${2\dot{\lambda_{1}}(y\dot{z}-z\dot{y})+2\dot{\lambda_{2}}(z\dot{x}-x\dot{z})\over
{x^2+y^2+z^2}} $ is a constant. Note that if $k=0$, both
$\ddot{x}$ and $\ddot{y}$ must be $0$, then $x$ and $y$ are both
linear function of t. $\lambda_{1}$ and $\lambda_{2}$ can be
easily obtained from (\ref{l1}) and (\ref{l2}).For the general
case the expressions involved are very complicated. Instead of
doing this we will present a particular solution.
 If we assume that $c_{2}=\frac{\sqrt{2}}{2}$, $c_{3}=1$,
$c_{6}=\frac{\sqrt{2}}{2}$ and others are $0$ , we obtain
\be\label{wr} x(t)=\frac{\sqrt{2}}{2}\cos(\sqrt{k}t) ,\quad
y(t)=sin(\sqrt{k}t)\quad and \quad
z(t)=\frac{\sqrt{2}}{2}\cos(\sqrt{k}t). \ee Using (\ref{wr}),
(\ref{l1}) and (\ref{l2}) we obtain
\begin{eqnarray}\label{sol}
  {\ddot\lambda_{1}}(t) &=& 0 \\
{\ddot\lambda_{1}}(t) &=&
\sqrt{2k}{\dot\lambda_{1}}(t)-\frac{2sin(\sqrt{k}t)\sqrt{k}}{\cos(\sqrt{k}t)}{\dot\lambda_{2}}(t).
\end{eqnarray}

The solution of (\ref{sol}) is
\be
\begin{array}{c}
\lambda_{1}(t)= k_{3}t+k_{4}, \\
 \lambda_{2}(t)=
\frac{1}{4}k_{3}ln(1+tan^2(\sqrt{k}t))\sqrt{2}\sqrt{\frac{1}{k}}+\frac{1}{2}k_{3}ln(\cos(\sqrt{k}t))\sqrt{2}\sqrt{\frac{1}{k}}\\
+
\frac{1}{4\sqrt{k}}k_{1}sin(2\sqrt{k}t)+(\frac{\sqrt{2}}{4\sqrt{k}}-\frac{\sqrt{2}}{4k}\cos(2\sqrt{k}t))k_{3}+k_{2}+\frac{1}{2}k_{1}t
\end{array}
\ee
 where $k_{i},i=1,\cdots 4$ are constants. As before $p_{i}$ have the forms $p_i=a_{ij}\dot{q^{j}}$.

\section{The singular case}

As an example of singular quadratic system, let's consider the
Lagrangian  given by
\begin{equation}\label{treii}
L=\frac{1}{2}(\dot{x^{2}}+\dot{y^{2}}+\dot{z^{2}})+\dot{\lambda_{1}}(y\dot{z}-z\dot{y})+\dot{\lambda_{2}}(z\dot{x}-x\dot{z})+\dot{\lambda_{3}}(x\dot{y}-y\dot{x})
\end{equation}
which is expressed in compact form as $L=\frac{1}{2}a_{ij}{\dot
q^{i}}{\dot q^{j}}$, where the matrix $a_{ij}$ has the form
\be\label{sase} a_{ij}=\left(%
\begin{array}{cccccc}\label{saispe}
  1 & 0 & 0 & 0 & z &-y\\
  0 & 1 & 0 & -z & 0& x\\
  0 & 0 & 1 & y & -x& 0\\
  0 & -z & y & 0 & 0& 0\\
  z & 0 & -x & 0 &0 & 0\\
 -y & x & 0  & 0 & 0& 0\\
\end{array}
\right). \ee

Since the rank of the matrix is 5, this is a singular system. By
using $p_{i}=\frac{\partial L}{\partial \dot{q^{i}}}$ we find
\begin{eqnarray}\label{uur}
p_{x}+y\dot{\lambda_{3}}=\dot{x}+z\dot{\lambda_{2}}, \quad
p_{y}-x\dot{\lambda_{3}}=\dot{y}-z\dot{\lambda_{1}},
p_{z}=a_{3i}\dot{q}^{i}, \quad p_{\lambda_{1}}=a_{4i}\dot{q}^{i},
\quad p_{\lambda_{2}}=a_{4i}\dot{q}^{i}. \end{eqnarray} From
(\ref{uur}) we find
\begin{equation}\label{uty}
\begin{array}{c}
  {\dot{x}}=\frac{p_{x}zx^2+p_{\lambda_{2}}z^2+p_{\lambda_{2}}y^2+xp_{z}z^2+
xyp_{\lambda_{1}}+xyzp_{y}
}{z(x^2+y^2+z^2)}, \\
  {\dot{y}}=\frac{p_{y}zy^2+p_{z}z^2y+p_{\lambda_{1}}z^2+xyzp_{x}-
xyp_{\lambda_{2}}-x^2p_{\lambda_{1}} }{z(x^2+y^2+z^2)},\\
  {\dot{z}}=\frac{p_{z}z^2+p_{y}yz+p_{\lambda_{1}}y-
x{\lambda_{2}}-xzp_{x}}{z(x^2+y^2+z^2)}, \\
 \dot{\lambda_{1}}=\frac{-p_{y}z^3+z^3x\dot{\lambda_{3}}+
 p_{z}z^2y-p_{\lambda_{1}}z^2+xyzp_{x}+x^3z\dot{\lambda_{3}}
-zx^2p_{y}+zx\dot{\lambda_{3}}y^2-
xyp_{\lambda_{2}}-x^2p_{\lambda_{1}}}{z(x^2+y^2+z^2)},  \\
\dot{\lambda_{2}}=\frac{-p_{\lambda_{2}}z^2-p_{\lambda_{2}}y^2
 z^3p_{x}+zy^2p_{x}+z^3y\dot{\lambda_{3}}+y^3z\dot{\lambda_{3}}-
p_{z}xz^2-p_{\lambda_{1}}xy-
p_{y}zyx+x^2yz\dot{\lambda_{3}}}{z(x^2+y^2+z^2)}.
\end{array}
\end{equation}
 Using (\ref{uty}), the canonical Hamiltonian has the following
 form
\begin{equation}
H_c=\frac{1}{2}b^{-1}_{\mu\rho}p_{\mu}p_{\rho}+\dot{\lambda_{3}}(p_{\lambda_{3}}+\frac{yp_{\lambda_{2}}+xp_{\lambda_{1}}}{z})
\quad where \quad \mu,\rho=1,..,5
\end{equation}
where $b^{-1}_{\mu\rho}$ is the inverse of \be\label{sapte}
 b_{\mu\rho}=\left(%
\begin{array}{ccccc}
  1 & 0 & 0 & 0 & z \\
  0 & 1 & 0 & -z & 0 \\
  0 & 0 & 1 & y & -x \\
  0 & -z & y & 0 & 0 \\
  z & 0 & -x & 0 &0 \\
\end{array}\right)
\ee In (HJ) formalism  the Hamiltonians to start with are

\begin{equation}\label{paa}
H_{0}^{'}=p_0 + \frac{1}{2}b^{-1}_{\mu\rho}p_{\mu}p_{\rho},\quad
H_{1}^{'}=p_{\lambda_{3}}+\frac{yp_{\lambda_{2}}+xp_{\lambda_{1}}}{z}.
\end{equation}

 The total differential equation corresponding to (\ref{paa})are
\begin{eqnarray}
&dx&=b_{1\rho}^{-1}p_{\rho}dt,\quad dy=b_{2\rho}^{-1}p_{\rho}dt,
\quad dz=b_{3\rho}^{-1}p_{\rho}dt,\\
&d\lambda_{1}&=b_{4\rho}^{-1}p_{\rho}dt+\frac{x}{z}d\lambda_{3},\quad
d\lambda_{2}=b_{5\rho}^{-1}dt+\frac{y}{z}d\lambda_{3}\\
&dp_{\lambda_{1}}&=0,\quad dp_{\lambda_{2}}=0, \quad
dp_{x}=\frac{-1}{2}(\frac{\partial b_{\mu\rho}^{-1}}{\partial
x})p_{\mu}p_{\rho}dt-\frac{p_{\lambda_{1}}}{z}d\lambda_{3},\\
&dp_{y}&=\frac{-1}{2}(\frac{\partial b_{\mu\rho}^{-1}}{\partial
y})p_{\mu}p_{\rho}dt-\frac{p\lambda_{2}}{z}d\lambda_{3},\\
&dp_{z}&=\frac{-1}{2}(\frac{\partial b_{\mu\rho}^{-1}}{\partial
z})p_{\mu}p_{\rho}dt+(\frac{yp_{\lambda_{2}}+xp_{\lambda_{2}}}{z^2})d\lambda_{3}
\end{eqnarray}
The next step is to find the variations  of $H_{0}^{'}$ and
$H_{1}^{'}$. Using the fact that $b^{-1}_{\mu\rho}$ is a symmetric
matrix and taking into account the previous equations we obtain
after some calculations that both variations are zero. We conclude
that the system is integrable and we observe that $H_1^{'}=z{\bf
{\vec r}}x {{\bf \vec L}}$ which is identically zero. The
geometrical properties of the $5X5$ metrics are presented in
Annex.
\subsection{Curved space generalization}

 Let us start with a free three dimensional Lagrangian
\be\label{lkl}
L=\frac{1}{2}(\dot{x^{2}}+\dot{y^{2}}+\dot{z^{2}})\ee If we impose
$z^2=1-x^2-y^2 $ and take its time derivative, we obtain \\
$\dot{z}=-\frac{x\dot{x}+y\dot{y}}{\sqrt{u}}$. So, (\ref{lkl})
becomes \be\label{ooof} L^{'}=g_{ab}{\dot q^a}{\dot q^b},\ee where
\be g_{ab}=\delta_{ab}+\frac{q^{a}q^{b}}{u}. \ee where $z^2=u,
 \quad a,b=1,2$.
The above metric admits three invariants \cite{nambum} \be
L_x=-\sqrt{u}p_y, L_y=\sqrt{u}p_x, L_z=xp_y-yp_x, \ee
 Adding those three constants of motion to
(\ref{ooof} ) we obtain the extended Lagrangian in a compact form
as $ L^{'}={1\over 2}a_{ij}{\dot q^i}{\dot q^j}$, where $a_{ij}$
is has the following form \be
 a_{ij}=\left(%
\begin{array}{ccccc}
1+\frac{x^2}{u} &\frac{xy}{u}  & -\frac{xy}{\sqrt{u}} &\frac{x^2}{\sqrt{u}}+\sqrt{u}  & -y \\
\frac{xy}{u}&  1+\frac{y^2}{u}& -\frac{y^2}{\sqrt{u}}-\sqrt{u} &\frac{xy}{\sqrt{u}}  & x \\
-\frac{xy}{\sqrt{u}}&-\frac{y^2}{\sqrt{u}}-\sqrt{u} &0  &0  & 0 \\
  \frac{x^2}{\sqrt{u}}+\sqrt{u} &\frac{xy}{\sqrt{u}} & 0 &0  & 0 \\
  -y & x &0  &0 & 0 \\
\end{array}%
\right). \ee The determinant $ a_{ij}$ is $0$ so $ a_{ij}$ is
 singular symmetric matrix and the rank of it is $4$.
In order to find the canonical Hamiltonian we need $\dot{x}$,
$\dot{y}$, $\dot{\lambda_{1}}$ and $\dot{\lambda_{2}}$ in terms of
$p_{x}$, $p_{y}$, $p_{\lambda_{1}}$, $p_{\lambda_{2}}$ and
 $\dot{\lambda_{3}}$ .Using
 \begin{equation}
p_{i}=\frac{\partial L}{\partial \dot{q^{i}}}=a_{ij}\dot{q^{j}}
\quad where \quad i,j=1,..,5,
\end{equation}
we obtain
 \be\begin{array}{c}
  \dot{x} =\frac{(-p_{\lambda_{2}}+p_{\lambda_{2}}x^2+xyp_{\lambda_{1}})\sqrt{1-x^2-y^2}}{1-x^2-y^2}\\
  \dot{y} =\frac{(p_{\lambda_{1}}-p_{\lambda_{1}}y^2+xyp_{\lambda_{2}})\sqrt{1-x^2-y^2}}{1-x^2-y^2}  \\
\dot{\lambda_{1}}=\frac{x^3\dot{\lambda_{3}}+x^3yp_{x}-x^2p_{y}+x^2p_{y}y^2+xy^2\dot{\lambda_{3}}-x\dot{\lambda_{3}}+xy\sqrt{1-x^2-y^2}+xy^3p_{x}}{\sqrt{1-x^2-y^2}(1-x^2-y^2)}\\
-\frac{xyp_{x}-p_{\lambda_{1}}\sqrt{1-x^2-y^2}y^2-2p_{y}y^2+p_{\lambda_{1}}\sqrt{1-x^2-y^2}+y^4p_{y}+p_{y}}{\sqrt{1-x^2-y^2}(1-x^2-y^2)} \\
\dot{\lambda_{2}}=\frac{p_{\lambda_{2}}\sqrt{1-x^2-y^2}(x^2-1)-p_{x}+2x^2p_{x}-y\dot{\lambda_{3}}+x^2y\dot{\lambda_{3}}-x^4p_{x}+y^2p_{x}-y^2x^2p_{x}}{\sqrt{1-x^2-y^2}(1-x^2-y^2)}\\
+\frac{y^3\dot{\lambda_{3}}-xy^3p_{y}+xyp_{\lambda_{1}}\sqrt{1-x^2-y^2-x^3yp_{y}+xyp_{y}}}{\sqrt{1-x^2-y^2}(1-x^2-y^2)}
\end{array}
\ee so, the canonical Hamiltonian is given by
\begin{equation}
H_c=\frac{1}{2}g^{-1}_{\mu\rho}p_{\mu}p_{\rho}+\dot{\lambda_{3}}(p_{\lambda_{3}}+\frac{yp_{\lambda_{2}}+xp_{\lambda_{1}}}{\sqrt{1-x^2-y^2}})
\quad where \quad \mu,\rho=1,..,4
\end{equation}
where
\be\label{nouaa} g_{\mu\rho}=\left(%
\begin{array}{cccc}
1+\frac{x^2}{u} &\frac{xy}{u}  & -\frac{xy}{\sqrt{u}} &\frac{x^2}{\sqrt{u}}+\sqrt{u} \\
\frac{xy}{u}&  1+\frac{y^2}{u}& -\frac{y^2}{\sqrt{u}}-\sqrt{u} &\frac{xy}{\sqrt{u}} \\
-\frac{xy}{\sqrt{u}}&-\frac{y^2}{\sqrt{u}}-\sqrt{u} &0  &0   \\
  \frac{x^2}{\sqrt{u}}+\sqrt{u} &\frac{xy}{\sqrt{u}} & 0 &0  \\
\end{array}%
\right). \ee
 Since the matrix $g_{\mu\rho}$ is singular, we obtain a constrained
system. In (HJ) formalism we have two Hamiltonians
\begin{equation}\label{sham}
H_{0}^{'}=p_{0}+\frac{1}{2}g^{-1}_{\mu\rho}p_{\mu}p_{\rho},\quad
H_{1}^{'}=p_{\lambda_{3}}+\frac{yp_{\lambda_{2}}+xp_{\lambda_{1}}}{\sqrt{1-x^2-y^2}}
\end{equation}

 The total differential equations of (\ref{sham}) are
given by
\begin{equation}\label{kkk}
\begin{array}{c}
dx=g_{1\rho}^{-1}p_{\rho}dt,\quad dy =
g_{2\rho}^{-1}p_{\rho}dt,\quad d\lambda_{1}=
g_{3\rho}^{-1}dt+\frac{x}{\sqrt{1-x^2-y^2}}d\lambda_{3},\\
d\lambda_{2}=g_{4\rho}^{-1}dt+\frac{y}{\sqrt{1-x^2-y^2}}d\lambda_{3},\\
dp_{x}= \frac{-1}{2}(\frac{\partial g_{\mu\rho}^{-1}}{\partial
x})p_{\mu}p_{\rho}dt-(\frac{p\lambda_{1}}{\sqrt{1-x^2-y^2}}+\frac{(xp_{\lambda_{1}}+yp_{\lambda_{2}})x}{(1-x^2-y^2)^{\frac{3}{2}}})d\lambda_{3}\\
dp_{y}= \frac{-1}{2}(\frac{\partial
g_{\mu\rho}^{-1}}{\partial y})p_{\mu}p_{\rho}dt-(\frac{p\lambda_{2}}{\sqrt{1-x^2-y^2}}+\frac{(xp_{\lambda_{1}}+yp_{\lambda_{2}})y}{(1-x^2-y^2)^{\frac{3}{2}}})d\lambda_{3}\\
dp_{\lambda_{1}}= 0, dp_{\lambda_{2}}=  0
\end{array}
\end{equation}
The variations of $H_0^{'}$ and $H_1^{'}$ are $0$. Geodesic
equations corresponding to $g_{\mu\rho}$ are given by
\begin{eqnarray}\label{muratino}
  \ddot{y}u &=&y(y^2-1)\dot{x}^2-2xy^2\dot{x}\dot{y}+y(x^2-1)\dot{y}^2  \\
 \ddot{x}u &=&x(y^2-1)\dot{x}^2-2yx^2\dot{x}\dot{y}+x(x^2-1)\dot{y}^2   \\
 \ddot{\lambda_{1}}u &=&-2x(y^2-1)\dot{x}\dot{\lambda_{1}}-2y(y^2-1)\dot{x}\dot{\lambda_{2}}+2x^2y\dot{y}\dot{\lambda_{1}}+2xy^2\dot{y}\dot{\lambda_{2}}  \\
 \ddot{\lambda_{2}}u &=&-2x(x^2-1)\dot{y}\dot{\lambda_{1}}-2y(x^2-1)\dot{y}\dot{\lambda_{2}}+2x^2y\dot{x}\dot{\lambda_{1}}+2xy^2\dot{x}\dot{\lambda_{2}}
\end{eqnarray}
where $u=1-x^2-y^2$. After some calculations  we obtain a solution

as \be \begin{array}{c} x(t)=cos(kt), y(t)=sin(kt), \\
\lambda_1(t) = {1\over k}e^{-{1\over 2}\pi+k t}\cos(k
t)C_4-{1\over k}\cos(k t)C_3 e^{(-1/2\pi-k
t)}+C_1, \\
\lambda_2(t) = (C_2k-C_3 e^{-{1\over 2}\pi-k t}\sin(k t)+{C_4
e^{(-{1\over 2}\pi+k t)}\sin(kt)\over k}.
\end{array}
 \ee

The geometrical properties of the corresponding metrics are
presented in Annex.
\section{Conclusions}
In this paper we illustrated the importance of the surface terms
in finding some integrable geometries in three, four and five
dimensions. The central role was played by the components of the
angular momentum. We added a surface term ${\dot\lambda_3}L_z$ to
a free two-dimensional Lagrangian and we obtain a three
dimensional geometry which is conformal flat and has a
non-vanishing scalar curvature. We solved the geodesics equations
and we found non-trivial solutions. We repeated the procedure and
we investigated the case when two components of the angular
momentum were added to a free three-dimensional Lagrangian.In this
manner we obtain a five-dimensional metric admitting three Killing
vectors. A singular system  was obtained if  all components of the
angular momentum were added. We treated this system within (HJ)
formalism and we obtained three types of four dimensional metrics
admitting three isometries. The procedure was extended to the
curved space and the two-dimensional sphere  was investigated in
details.

 \section {Acknowledgments}
One of the authors (D. B.) would like to thank T. L. Curtright for
valuable comments and remarks. This work is partially supported by
the Scientific and Technical Research Council of Turkey.

\section{Annex}
 In the following  the Christoffel components,
Ricci\quad scalar and the Weyl components of the obtained metrics
are presented.

{\bf 1.} For the singular case we obtain three $5x5$ non-singular
matrices from (\ref{sase}).Let us denote k as
$k=x^2+y^2+z^2$.The first metric is given by (\ref{sapte}).\\
 The non-zero Christoffel components are given
$$\Gamma^{1}_{24}=-\Gamma^{1}_{15}=-\frac{z}{y}\Gamma^{1}_{34}=\frac{z}{x}\Gamma^{1}_{35}=\Gamma^{1}_{24}
-\frac{x}{y}\Gamma^{2}_{15}=\frac{x}{y}\Gamma^{2}_{24}=-\frac{zx}{y^2}\Gamma^{2}_{34}=\frac{z}{y}\Gamma^{2}_{35}$$
$$=-\frac{x}{z}\Gamma^{3}_{15}=\frac{x}{z}\Gamma^{3}_{24}=-\frac{x}{y}\Gamma^{3}_{34}=
\Gamma^{3}_{35}=-\frac{zx}{y}\Gamma^{4}_{15}=\frac{zx}{y}\Gamma^{4}_{24}
=\frac{z^2x}{x^2+z^2}\Gamma^{4}_{34}=\frac{z^2}{y}\Gamma^{4}_{35}$$
$$=z\Gamma^{5}_{15}=-z\Gamma^{5}_{24}=\frac{z}{y}\Gamma^{5}_{34}=\frac{z^2x}{z^2+y^2}\Gamma^{5}_{35}=\frac{zx}{k}$$

Ricci\quad scalar is given by $R=\frac{12}{x^2+y^2+z^2}$ and
 Weyl tensor components are\\
$$W_{1212}=-\frac{1}{3}\frac{2x^2+z^2+2y^2}{k^2},W_{1213}=-\frac{1}{3}\frac{yz}{k^2},
W_{1214}=-\frac{1}{3}\frac{(z^2+y^2)z}{k^2},W_{1423}=-\frac{1}{3}\frac{(z^2+y^2)x}{k^2}$$
$$W_{1215}=W_{1224}=\frac{1}{3}\frac{xyz}{k^2},W_{1223}=\frac{1}{3}\frac{xz}{k^2}
,W_{1225}=-\frac{1}{3}\frac{z(x^2+z^2)}{k^2},W_{2335}=\frac{1}{3}\frac{xyz}{k^2}$$
$$W_{1234}=\frac{1}{3}\frac{z^2x}{k^2},W_{1234}=\frac{1}{3}\frac{yz^2}{k^2},W_{1313}=-\frac{1}{3}\frac{y^2+2x^2+2z^2}{k^2},W_{2323}=-\frac{1}{3}\frac{2y^2+x^2+2z^2}{k^2}$$
$$W_{1523}=W_{2324}=\frac{1}{3}\frac{x^2y}{k^2},W_{1314}=\frac{1}{3}\frac{(z^2+y^2)y}{k^2},W_{1315}=W_{1324}=-\frac{1}{3}\frac{xy^2}{k^2},W_{1323}=-\frac{1}{3}\frac{xy}{k^2} $$
$$W_{1325}=\frac{1}{3}\frac{(z^2+x^2)y}{k^2},W_{1334}=-\frac{1}{3}\frac{xyz}{k^2},
W_{1335}=-\frac{1}{3}\frac{y^2z}{k^2},W_{2334}=\frac{1}{3}\frac{x^2z}{k^2},W_{2325}=-\frac{1}{3}\frac{(x^2+z^2)x}{k^2}.$$

 The second metric is given by

\be b_{\mu\nu}^{(2)}=\left(%
\begin{array}{ccccc}
  1 & 0 & 0 & 0 & -y \\
  0 & 1 & 0 & -z & x \\
  0 & 0 & 1 & y & 0 \\
  0 & -z & y & 0 & 0 \\
  -y & x & 0 & 0 &0 \\
\end{array}%
\right) \ee

The non-zero Christoffel symbols are given by
$$\Gamma^{1}_{15}=\frac{y}{z}\Gamma^{1}_{24}=-\frac{y}{x}\Gamma^{1}_{25}=-\Gamma^{1}_{34}=\frac{x}{y}\Gamma^{2}_{15}
=\frac{x}{z}\Gamma^{2}_{24}=-\Gamma^{2}_{25}=-\frac{x}{y}\Gamma^{2}_{34}=\frac{x}{z}\Gamma^{3}_{34}$$
$$\frac{x}{z}\Gamma^{3}_{15}=\frac{xy}{z^2}\Gamma^{3}_{24}=-\frac{y}{z}\Gamma^{3}_{25}=-\frac{xy}{z}\Gamma^{4}_{15}
=\frac{xy^2}{x+y}\Gamma^{4}_{24}=\frac{y^2}{z}\Gamma^{4}_{25}=\frac{xy}{z}\Gamma^{4}_{34}=y\Gamma^{5}_{15}$$
$$\frac{y^2}{z}\Gamma^{5}_{24}=\frac{xy^2}{z^2+y^2}\Gamma^{5}_{25}=-y\Gamma^{5}_{35}==-\frac{x}{z}\Gamma^{3}_{34}=\frac{xy}{k}$$
$$Ricci\quad scalar\quad R=\frac{12}{x^2+y^2+z^2}.$$
The Weyl tensor components are\\
$$W_{1212}=-\frac{1}{3}\frac{z^2+2x^2+2y^2}{k^2},W_{1213}=-\frac{1}{3}\frac{zy}{k^2},W_{1214}=-\frac{1}{3}\frac{(z^2+y^2)z}{k^2},W_{1215}=\frac{1}{3}\frac{z^2x}{k^2}$$
$$W_{1223}=\frac{1}{3}\frac{zx}{k^2},W_{1224}=\frac{1}{3}\frac{xyz}{k^2},W_{1225}=\frac{1}{3}\frac{z^2y}{k^2},W_{1234}=\frac{1}{3}\frac{z^2x}{k^2},W_{1235}=-\frac{1}{3}\frac{(x^2+y^2)z}{k^2}$$
$$W_{1313}=-\frac{1}{3}\frac{2z^2+2x^2+y^2}{k^2},W_{1314}=\frac{1}{3}\frac{y(z^2+y^2)}{k^2},W_{1315}=-\frac{1}{3}\frac{xyz}{k^2},W_{1323}=-\frac{1}{3}\frac{xy}{k^2}$$
$$W_{1324}=-\frac{1}{3}\frac{y^2x}{k^2},W_{1325}=-\frac{1}{3}\frac{y^2z}{k^2},W_{1334}=-\frac{1}{3}\frac{xyz}{k^2},W_{1335}=\frac{1}{3}\frac{(x^2+y^2)y}{k^2}$$
$$W_{1423}=-\frac{1}{3}\frac{(z^2+y^2)x}{k^2},W_{1523}=\frac{1}{3}\frac{zx^2}{k^2},W_{2323}=-\frac{1}{3}\frac{2z^2+2y^2+x^2}{k^2},W_{2324}=-\frac{1}{3}\frac{x^2y}{k^2}$$
$$W_{2325}=\frac{1}{3}\frac{xyz}{k^2},W_{2334}=\frac{1}{3}\frac{zx^2}{k^2},W_{2335}=-\frac{1}{3}\frac{(x^2+y^2)x}{k^2}$$

\be
 b_{\mu\nu}^{(3)}=\left(%
\begin{array}{ccccc}
  1 & 0 & 0 &z & -y \\
  0 & 1 & 0 & 0 & x \\
  0 & 0 & 1 & -x & 0 \\
  z & 0 & -x & 0 & 0 \\
  -y & x & 0 & 0 &0 \\
\end{array}
\right) \ee The non-zero Christoffel symbols of the second kind
are given as

$$\Gamma^{1}_{14}=-\frac{z}{y}\Gamma^{1}_{15}=\frac{z}{x}\Gamma^{1}_{25}=-\frac{z}{x}\Gamma^{1}_{34}=\frac{x}{y}\Gamma^{2}_{14}
=-\frac{zx}{y^2}\Gamma^{2}_{15}=\frac{z}{y}\Gamma^{2}_{25}=-\frac{z}{y}\Gamma^{2}_{34}=\frac{x}{z}\Gamma^{3}_{14}$$
$$-\frac{x}{y}\Gamma^{3}_{15}=\Gamma^{3}_{25}=-\Gamma^{3}_{34}=-\frac{zx}{x^2+y^2}\Gamma^{4}_{14}
=-\frac{x}{y}\Gamma^{4}_{15}=x\Gamma^{4}_{25}=-x\Gamma^{4}_{34}=-\frac{x}{y}\Gamma^{5}_{14}$$
$$=-\frac{zx^2}{x^2+z^2}\Gamma^{5}_{15}=-\frac{zx}{y}\Gamma^{5}_{25}=\frac{zx}{y}\Gamma^{5}_{34}=-\frac{zx}{k}$$

Ricci\quad scalar R=$\frac{12}{x^2+y^2+z^2}$, Weyl tensor
components are
$$W_{1212}=-\frac{1}{3}\frac{z^2+2x^2+2y^2}{k^2},W_{1213}=-\frac{1}{3}\frac{zy}{k^2},W_{1214}=\frac{1}{3}\frac{zyx}{k^2},W_{1215}=\frac{1}{3}\frac{z^2x}{k^2}$$
$$W_{1223}=\frac{1}{3}\frac{zx}{k^2},W_{1224}=\frac{1}{3}\frac{(z^2+x^2)z}{k^2},W_{1225}=W_{1234}=\frac{1}{3}\frac{z^2y}{k^2},W_{1235}=-\frac{1}{3}\frac{(x^2+y^2)z}{k^2}$$
$$W_{1313}=-\frac{1}{3}\frac{2z^2+2x^2+y^2}{k^2},W_{1314}=-\frac{1}{3}\frac{y^2x}{k^2},W_{1315}=-\frac{1}{3}\frac{xyz}{k^2},W_{1323}=-\frac{1}{3}\frac{xy}{k^2}$$
$$W_{1324}=\frac{1}{3}\frac{(x^2+z^2)y}{k^2},W_{1325}=-\frac{1}{3}\frac{y^2z}{k^2},W_{1334}=-\frac{1}{3}\frac{zy^2}{k^2},W_{1335}=\frac{1}{3}\frac{(x^2+y^2)y}{k^2}$$
$$W_{1423}=\frac{1}{3}\frac{x^2y}{k^2},W_{1523}=\frac{1}{3}\frac{zx^2}{k^2},W_{2323}=-\frac{1}{3}\frac{2z^2+2y^2+x^2}{k^2},W_{2324}=-\frac{1}{3}\frac{(x^2+z^2)x}{k^2}$$
$$W_{2325}=\frac{1}{3}\frac{xyz}{k^2},W_{2334}=\frac{1}{3}\frac{xyz}{k^2},W_{2335}=-\frac{1}{3}\frac{(x^2+y^2)x}{k^2}.$$

 {\bf 2.}
 We are repeating the same procedure as above for the curved case.
 The metric given (\ref{nouaa}) has Ricci\quad scalar is 0,$W_{1212} = -\frac{1}{u}$, Christoffel symbols of
the second kind non-zero components
$$\Gamma^{1}_{11}=-\Gamma^{3}_{13}=\frac{x(1-y^2)}{u},\quad
\Gamma^{2}_{22}=-\Gamma^{4}_{24}= \frac{y(1-x^2)}{u},\quad
\Gamma^{1}_{22}=-\Gamma^{4}_{23}=\frac{x(1-x^2)}{u}$$
$$\Gamma^{4}_{14}=\Gamma^{3}_{24}=-\Gamma^{2}_{12}=-\frac{xy^2}{u},
\quad
-\Gamma^{1}_{12}=\Gamma^{3}_{23}=\Gamma^{4}_{13}=-\frac{x^2y}{u},
\Gamma^{2}_{11}=-\Gamma^{3}_{14}=\frac{y(1-y^2)}{u},$$
$u=1-x^2-y^2$.

\be\label{fgf}
g_{\mu\nu}^{(2)}=\left(%
\begin{array}{cccc}
1+\frac{x^{2}}{u}&\frac{xy}{u}&-\frac{xy}{\sqrt{u}}&-y\\
\frac{xy}{u}&1+\frac{y^2}{u}&-\sqrt{u}-\frac{y^2}{\sqrt{u}}&x\\
-\frac {xy}{\sqrt{u}}&-\sqrt{u}-\frac {y^2}{\sqrt{u}} & 0&0   \\
-y&x& 0 &0   \\
\end{array}%
\right) \ee
 Ricci\quad scalar is R=0,$W_{1212} = -\frac{1}{u}$

The Christoffel symbols are
$$\Gamma^{1}_{11}=
\frac{x(1-y^2)}{u},\Gamma^{1}_{12}=\frac{x^2y}{u},\Gamma^{1}_{22}=\frac{x(1-x^2)}{u},
\Gamma^{2}_{11}=\frac{y(1-y^2)}{u},\Gamma^{2}_{12}=\frac{xy^2}{u},\Gamma^{2}_{22}=
\frac{y(1-x^2)}{u} $$
$$\Gamma^{3}_{13}=-x,\Gamma^{3}_{14}=-\sqrt{u},\Gamma^{3}_{23}=-\frac{x^2}{y},
\Gamma^{3}_{24}=\frac{x}{y}\sqrt{u},\Gamma^{4}_{13}=-\frac{x^2}{\sqrt{u}},
\Gamma^{4}_{14}=x$$$$
\Gamma^{4}_{23}=\frac{x(-1+x^2)}{\sqrt{-u}},\Gamma^{4}_{24}=\frac{x^2-1}{y}$$
\be\label{hgh}
g_{\mu\nu}^{(3)}=\left(%
\begin{array}{cccc}
1+\frac{x^{2}}{u}&\frac{xy}{u}&\sqrt{u}+\frac{x^2}{\sqrt{u}}&-y\\
\frac{xy}{u}&1+\frac{y^2}{u}&\frac{xy}{\sqrt{u}}&x\\
\sqrt{u}+\frac{x^2}{\sqrt{u}}&\frac{xy}{\sqrt{u}} & 0&0   \\
-y&x& 0 &0   \\
\end{array}%
\right) \ee

Ricci\quad scalar is R=0, $W_{1212}= -\frac{1}{u}$.
 The Christoffel symbols are
$$\Gamma^{1}_{11}=
\frac{x(1-y^2)}{u},\Gamma^{1}_{12}=\frac{x^2y}{u},\Gamma^{1}_{22}=\frac{x(1-x^2)}{u},
\Gamma^{2}_{11}=\frac{y(1-y^2)}{u},\Gamma^{2}_{12}=\frac{xy^2}{u}$$
$$\Gamma^{2}_{22}=\frac{y(1-x^2)}{u}, \Gamma^{3}_{13}=\frac{y^2}{x},\Gamma^{3}_{14}=\frac{y}{x}\sqrt{u}
,\Gamma^{3}_{23}=-y,\Gamma^{3}_{24}=-\sqrt{u}$$
$$\Gamma^{4}_{13}=\frac{(-y^2+1)y}{x\sqrt{-u}},\Gamma^{4}_{14}=\frac{1-y^2}{x},
\Gamma^{4}_{23}=\frac{y^2}{\sqrt{u}},\Gamma^{4}_{24}=y$$

\end{document}